\documentclass[onecolumn]{article}

\usepackage{srcltx}

\usepackage{cite}
\usepackage{url}
\usepackage{amsmath,amssymb,amsfonts}
\usepackage{mathdots}
\usepackage{epsfig}
\usepackage{color}
\usepackage{flushend}

\usepackage{graphics}
\usepackage{graphicx}
\usepackage{subfigure}
\usepackage{hyperref}

\usepackage{fullpage}
\usepackage{setspace}
\usepackage{booktabs}

\usepackage{mathptmx}

\usepackage{multirow}

\title{Multiplierless Approximate 4-point DCT VLSI 
Architectures for Transform Block Coding}

\author{F.~M.~Bayer%
\thanks{F. M. Bayer 
is with the
Departamento de Estat\'istica and 
Laborat\'orio de Ci\^encias Espaciais de Santa Maria (LACESM),
Universidade Federal de Santa Maria, RS, Brazil,
E-mail: bayer@ufsm.br}
\and
R.~J.~Cintra%
\thanks{R. J. Cintra 
is with
the Signal Processing Group, 
Departamento de Estat\'istica, 
Universidade Federal de Pernambuco, PE, Brazil,
E-mail: rjdsc@stat.ufpe.org}
\and
A.~Madanayake%
\thanks{A.~Madanayake and U.~S.~Potluri 
are with the
ECE, The University of Akron, Akron, OH, USA,
E-mail: arjuna@uakron.edu}
\and
U.~S.~Potluri${}^\ddagger$}

\date{}

\begin{document}

\maketitle

\doublespacing

\abstract{Two multiplierless algorithms are proposed for 4$\times$4
approximate-DCT for transform coding in digital video. 
Computational architectures for \mbox{1-D}/\mbox{2-D} 
realisations are implemented using Xilinx FPGA devices.
CMOS synthesis at the 45~nm node indicate real-time operation at 
1~GHz yielding 4$\times$4 block rates of 125~MHz at
less than 120~mW of dynamic power consumption.}

\section{Introduction}
Video and multimedia processing based on signal and image compression such as the high efficiency video coding (HEVC) and H.265 reconfigurable video codecs require \mbox{2-D} transform block coding 
for block sizes $N\times N$ where $N\in\{4,8,16,32,64\}$ \cite{sullivan2012}. 
The transform coding stage requires algorithms
for the $N$-point discrete cosine transform (DCT)
of types II and IV. 
The associate transformation matrices
are defined, respectively,
according to~\cite{britanak2007discrete}:
\begin{align*}
\left[\mathbf{C}_\text{II}\right]_{(m,n)}
=
\sqrt{\frac{2}{N}}
\cdot
\alpha_m 
\cdot
\cos 
\left[ 
\left( m - \frac{1}{2} 
\right) 
\cdot
\frac{\pi(n-1)}{N} 
\right]
,
\\
\left[\mathbf{C}_\text{IV}\right]_{(m,n)}
=
\sqrt{\frac{2}{N}}
\cdot
\cos 
\left[ 
\left( m - \frac{1}{2} \right) 
\cdot
\left( n - \frac{1}{2} \right) 
\cdot
\frac{\pi}{N} 
\right]
,
\end{align*}
where $m,n=1,2,\ldots,N$,
$\alpha_1 = 1/\sqrt{2}$,
and 
$\alpha_m = 1$, for $m>1$.

In this letter,
our goal is to propose
multiplication-free approximations for the 4-point
DCT-II and -IV as well as its fast algorithms.
We also aim at  VLSI realisations
of both \mbox{1-D} and \mbox{2-D}
versions of the derived approximate transforms,
while
maintaining
at high numerical accuracy and low computational complexity.

\section{Optimisation and orthogonalization}%

Let $\mathcal{M}_P(4)$ be the set of all 4$\times$4 matrices whose entries
are defined over $P = \{-1, 0, 1 \}$.
In this set,
all matrices represent multiplierless transformations.
Our goal is to find matrices in $\mathcal{M}_P(4)$
that satisfactorily approximate
$\mathbf{C}_\text{II}$ 
and $\mathbf{C}_\text{IV}$.

Therefore
we propose the following multivariate non-linear 
optimisation problem over $\mathcal{M}_P(4)$
\begin{align}
\label{equation-optimization}
\mathbf{C}_k^\ast
&=
\arg 
\min_{\mathbf{A} \in \mathcal{M}_P(4)}
\operatorname{error}( \mathbf{A}, \mathbf{C}_k),
\quad
k\in\{\text{II}, \text{IV} \}
,
\end{align}
where 
$\mathbf{C}_k^\ast$ are the optimal matrices
and 
$\operatorname{error}(\cdot,\cdot)$
is an error measure between a given candidate matrix and
the exact matrices 
$\mathbf{C}_\text{II}$ 
and $\mathbf{C}_\text{IV}$.

Let $h_i[n]$ be the discrete signal formed by 
the $i$th row of a given matrix~$\mathbf{T}$
and the discrete-time Fourier transform (DTFT) of $h_i[n]$ 
be denoted by $H_i(\omega; \mathbf{T})$.
As discussed in~\cite{haweel2001new,cb2011}, 
we adopted the total error energy
as the error measure.
This particular measure is defined as follows:
\begin{align*}
\label{equation-total-error-energy}
\epsilon(\mathbf{A},\mathbf{C}_k)
=
\sum_{m=1}^4
\int_0^\pi
\left|
H_m(\omega; \mathbf{A})
-
H_m(\omega; \mathbf{C}_k)
\right|^2
\mathrm{d}\omega
,
\end{align*}
for $k\in\{\text{II}, \text{IV} \}$.
In other words,
$\epsilon(\mathbf{A},\mathbf{C}_k)$
quantifies 
the sum of the 
energy error in the DTFT domain---between~$\mathbf{A}$ and $\mathbf{C}_k$---when the entries of a given matrix row are
interpreted as filter coefficients~\cite{haweel2001new,cb2011}.
This quantity can be computed numerically    
by standard quadrature methods~\cite{piessens1983quadpack}.

As an additional
constraint to~\eqref{equation-optimization},
we impose that
the matrix $\mathbf{A}\cdot\mathbf{A}^{\top}$
must be a diagonal matrix
to ensure that orthogonality 
can be achieved in the obtained approximations~\cite{cintra2011integer}.
The resulting constrained optimisation problem
is algebraically intractable
and we resorted to exhaustive computational search.

\section{Proposed 4-point DCT approximations}
\label{s:proposed}

By solving~\eqref{equation-optimization},
we obtained
the following new DCT approximations:
\begin{align*}
\mathbf{C}^\ast_\text{II}
=
\left[
\begin{array}{rrrr}
 1 &  1 &  1 &  1 \\
 1 &  0 &  0 & -1 \\
 1 & -1 & -1 &  1 \\
 0 & -1 &  1 &  0 
\end{array}
\right]
\quad
\text{and}
\quad
\mathbf{C}^\ast_\text{IV}
=
\left[
\begin{array}{rrrr}
 1 &  1 &  1 &  0 \\
 1 &  0 & -1 & -1 \\
 1 &  -1 &  0 &  1 \\
 0 &  -1 &  1 & -1
\end{array}
\right]
.
\end{align*}
Although possessing very low complexity,
these matrices are not orthogonal.
In several contexts,
such as image processing for coding,
orthogonality is often a desirable property~\cite{britanak2007discrete}.
Adopting the orthogonalization methods
detailed in~\cite{cintra2011integer},
new orthogonal matrices
$\hat{\mathbf{C}}_\text{II}$
and
$\hat{\mathbf{C}}_\text{IV}$
can be derived
based on
$\mathbf{C}^\ast_\text{II}$
and
$\mathbf{C}^\ast_\text{IV}$,
respectively.
These orthogonal matrices are given by:
\begin{align*}
\hat{\mathbf{C}}_\text{II}
=
\mathbf{D}_\text{II} \cdot \mathbf{C}^\ast_\text{II}
\quad
\text{and}
\quad
\hat{\mathbf{C}}_\text{IV}
=
\mathbf{D}_\text{IV} \cdot \mathbf{C}^\ast_\text{IV}
,
\end{align*}
where 
$\mathbf{D}_\text{II}
= 
\sqrt{[
\mathbf{C}^\ast_\text{II} 
\cdot 
(\mathbf{C}^\ast_\text{II})^{\top}
]^{-1}
}
$
and
$\mathbf{D}_\text{IV}
=
\sqrt{
[
\mathbf{C}^\ast_\text{IV} 
\cdot (\mathbf{C}^\ast_\text{IV})^{\top}
]^{-1}
}
$.
Explicitly we obtain that
\begin{align*}
\mathbf{D}_\text{II}=
\operatorname{diag}
\left(
\frac{1}{2},\frac{1}{\sqrt{2}},\frac{1}{2},\frac{1}{\sqrt{2}}
\right)
\end{align*}
and
\begin{align*}
\mathbf{D}_\text{IV}=\frac{1}{\sqrt{3}} \cdot  \mathbf{I}_4
,
\end{align*}
where $\mathbf{I}_4$ is the identity matrix of size~4.
In image compression context 
the scaling matrices 
$\mathbf{D_1}$ and $\mathbf{D_2}$
may not introduce any computational overhead,
because they can be merged 
into the quantisation step,
as described earlier in~\cite{bas2008,cb2011,bc2012,bas2011}.

\begin{figure}
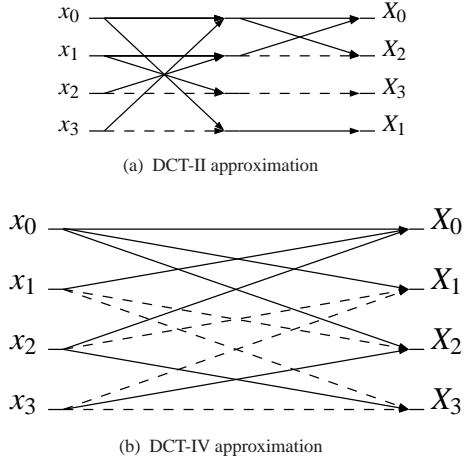

\centering
\subfigure[DCT-II approximation%
]{\scalebox{1} {\input{diagram_DCTII.tex}}} %
\\
\subfigure[DCT-IV approximation%
]{\scalebox{1}  {\input{diagram_DCTIV.tex}}} %
\caption{
Signal flow graph for proposed transforms.
}
\label{f:fast}
\end{figure}

The signal flow graph for 
$\mathbf{C}^\ast_{\text{II}}$
and  
$\mathbf{C}^\ast_{\text{IV}}$
is shown in Fig.~\ref{f:fast}.
We note that 
the $\mathbf{C}^\ast_{\text{II}}$ 
and $\mathbf{C}^\ast_{\text{IV}}$
transformations 
require only 6 and 8 additions,
respectively.
Multiplications or bit-shifting operations are totally absent.
Resulting approximations 
$\hat{\mathbf{C}}_\text{II}$ and $\hat{\mathbf{C}}_\text{IV}$ 
are very
close to the respective ideal DCT
and
offer extremely low complexities.
In Table~\ref{comparison},
we show 
the error measure 
and arithmetic complexity 
for
the proposed transforms, 
the exact DCT computation~\cite{britanak2007discrete},
and 
the well-known signed DCT~\cite{haweel2001new}.

\begin{table}%
\centering
\caption{Total error energy and arithmetic complexity analysis}
\label{comparison}
\begin{tabular}{lcc c c}
\toprule
\multirow{2}{1.5cm}{Method} & 
\multirow{2}{1.5cm}{\centering Error Energy} & 
\multicolumn{3}{c}{Complexity} \\
 &  & Add. &  Mult. & Total\\
\midrule
Exact 4-point DCT-II~\cite{britanak2007discrete} & 
0.000 & 8 & 4 & 12\\
4-point Signed DCT-II~\cite{haweel2001new}  & 
0.957 & 8  & 0 & 8\\ 
Proposed $\hat{\mathbf{C}}_\text{II}$ & 
0.957 & 6  & 0 & 6\\
\midrule
Exact 4-point DCT-IV~\cite{britanak2007discrete} & 
0.000 & 12 & 8 & 20\\ 
4-point Signed DCT-IV~\cite{haweel2001new} & 
2.359 & 10 &  0 & 10\\ 
Proposed $\hat{\mathbf{C}}_\text{IV}$ & 
0.838 & 8 &  0 & 8\\
\bottomrule
\end{tabular}
\end{table}

\section{FPGA prototypes}
\label{s:hardware}

The approximate DCTs were realised as an architecture 
for 
the 4-point \mbox{1-D} transforms 
and 
the extended to 4$\times$4 \mbox{2-D} transformation.
The inputs were assumed at 8-bit resolution.
Rapid prototypes were realised on a Xilinx Virtex-6
field programmable gate array (FPGA) device and tested to ensure correct on-chip
functionality. 
The results concerning the consumption
of
configurable logic blocks (CLB), flip-flops (FF), look-up tables (LUT),
and slices are shown in Table~\ref{fpga}. 
The maximum operating frequency ($F_\text{max}$) 
and dynamic power consumption ($D_p$)
are also displayed.

The register transfer language (RTL) code
corresponding to the FPGA-verified designs were targeted to 
45~nm CMOS standard cell
process using Cadence Encounter. 
The CMOS designs were realised up to synthesis and
place-and-route levels leading to the estimated results in Table~\ref{asic}.
Area-time complexities AT and AT${}^2$ were adopted
and measured 
in $\mathrm{\mu m^2 \cdot ns}$ and $\mathrm{\mu m^2 \cdot ns^2}$, respectively.

\begin{table}%
\centering
\caption{Resource consumption on Xilinx XC6VSX475T-2FF1156}
\label{fpga}
\begin{tabular}{l c c c c c c}
\toprule
\multirow{2}{1.5cm}{Proposed Approx.} & 
\multirow{2}{0.7cm}{CLB} & 
\multirow{2}{0.5cm}{\centering FF} & 
\multirow{2}{0.6cm}{\centering LUT} & 
\multirow{2}{0.7cm}{\centering Slices} &
\multirow{2}{0.8cm}{\centering $F_\text{max}$ (MHz)} &
\multirow{2}{0.7cm}{\centering $D_p$ (W)} 
\\
\\
\midrule 
1-D DCT-II & 56 & 76 &92 & 35 & 743.5 & 0.535\\
1-D DCT-IV & 76& 132 & 128 & 52  & 735.3 & 0.574 \\
2-D DCT-II & 166 & 408&330 & 108 & 704.2 & 0.884\\
2-D DCT-IV & 210& 528 & 472 & 148 & 689.2 & 0.921\\
\bottomrule
\end{tabular}
\end{table}

\begin{table}%
\centering
\caption{Resource consumption for 45~nm CMOS}
\label{asic}
\begin{tabular}{lc c c c c c }
\toprule
\multirow{2}{1.4cm}{Proposed Approx.} & 
\multirow{2}{0.7cm}{\centering ASIC Gates}
&
\multirow{2}{0.8cm}{\centering Area ($\mathrm{\mu m^2}$)} & 
\multirow{2}{0.8cm}{\centering $F_{\text{max}}$ (GHz)} & 
\multirow{2}{0.7cm}{\centering $D_p$ (mW)} &
\multirow{2}{0.5cm}{\centering AT} & 
\multirow{2}{0.5cm}{\centering AT${}^2$}
\\
\\
\midrule 
1-D DCT-II & 849 & 3386.9&1.10 &6.31 & 3160 & 2948\\
1-D DCT-IV & 1207& 4870.4 & 1.00 & 8.62 & 4846 & 4822 \\
2-D DCT-II & 7400 & 31217.8&0.95 &59.33 & 7770 & 8159\\
2-D DCT-IV & \!\!\!13770 & 59052.5&0.94 \!\!\! &\!\!\!115.66& \!\!\!14596 & \!\!\!15472\\
\bottomrule
\end{tabular}
\end{table}

\section{Conclusion}
\label{s:conclusion}

Numerical optimisation methods 
have lead to 4-point approximations for 
the DCT-II and DCT-IV.
Such
matrices are tailored for minimal computational complexity
and are adequate for computing 
realisations linked to
coding operations with applications in digital video and multimedia. 
Fast algorithms were derived and the associate physical realisations
do not require VLSI area- and power-intensive multiplier circuits.
Both 1-D and 2-D realisations were proposed 
with FPGA prototypes for architecture validation 
and 
CMOS synthesis results at the 45~nm node.
Results indicate real-time blockrate of 125~MHz for processing
4$\times$4 blocks at 1~GHz clock frequency.

\section*{Acknowledgments}

We thank The College of Engineering at UA, CNPq, FACEPE, and FAPERGS for 
the partial financial support.

\end{document}